# The structural effects of (111) growth of $La_2CoMnO_6$ on $SrTiO_3$ and LSAT – new insights from 3D crystallographic characterisation with 4D-STEM and Digital Dark Field imaging

Ian MacLaren[1], Andrew T. Fraser[1], Matthew R. Lipsett[1]

1. School of Physics and Astronomy, University of Glasgow, Glasgow G12 8QQ, UK

The 3-dimensional orientation of La atom modulations has been mapped in two thin films of $La_2CoMnO_6$ grown on $SrTiO_3$ and LSAT ([La,Sr,Al,Ta] oxide) using a 4D-scanning transmission electron microscopy (4D-STEM) method based on the recently developed Digital Dark Field method. This images the shifts of diffraction spots and the azimuthal intensity distribution in the First Order Laue Zone, and then uses them to reconstruct and map the 3D crystallography. This clearly shows a flip from out-of-plane modulation with tensile strain on $SrTiO_3$ to in-plane modulation with compressive strain on LSAT. This hitherto unobserved crystallographic change had a significant influence on the out-of-plane lattice parameter which left more room for the full incorporation of the larger $CoO_6$ octahedra in the film grown on LSAT and therefore explained the improved Mn-Co ordering and better properties for this film. Moreover, the method would be applicable to many other systems of epitaxial growth of complex oxides, revealing crystallographic details of crucial importance to properties which are not visible in conventional atomic resolution imaging.

## I. Introduction

Dark field imaging in the transmission electron microscope of domain orientations in functional oxides and thin films has been well used down the years [1-8], particularly to isolate specific domains based on the presence or absence of specific superlattice reflections [2,6,7]. Whilst similar information has often been obtained by atomic-resolution imaging techniques [9-14], dark field imaging has the advantage of being able to image much larger sample areas and thereby give much more statistically useful information and a better sense of the overall picture. In the case of thin epitaxial films, these may in turn reveal the crystal orientation and thus enable the characterisation of the preferred orientations (e.g. MacLaren *et al.* [2]), which can then be compared with arguments based on strain or other considerations to understand why the observed domain structure has formed. As a result, this can contribute to the optimisation of film properties by strain engineering, by growing on an appropriate substrate that gives the best domain orientations to achieve the desired properties.

In the last decade or so, dark field imaging has been extended to STEM (scanning transmission electron microscopy) using 4D-STEM techniques [15-17] (4-dimensional STEM – 2 dimensions of real space and 2 of reciprocal space) with small convergence angles such as SEND (scanning electron nanodiffraction), SPED (scanning precession electron diffraction) and other names for the same things. The most accepted current term for this imaging method [18,19] is *Virtual Dark Field* (VDF), although it is not entirely clear who invented the term. It has been found that arrays of apertures are especially powerful for this purpose [20-23]. Recently, we introduced a *Digital Dark Field* (DDF) method based on reducing 4D-STEM datasets to a sparse list of diffraction spots (and their intensities) [17,24] and only counting their intensity if they fit to one of the expected positions for a specific set of spots for a given crystal or domain orientation [25]. This has the advantage over the earlier VDF method that it only adds up intensity in discrete diffraction spots with centroids matching some specific expected position or set of positions, and thereby ignores nearby spots from other phases and general diffuse scattering. This therefore gives much greater specificity and contrast than traditional integration of all intensity with an aperture or array of apertures. In the current work, where much of the dark field imaging is carried out with weak superlattice or high angle diffraction spots, this improved selectivity and contrast from DDF is critical and the work would not have been possible using standard VDF imaging.

However, the vast majority of dark field imaging studies published to-date, either in TEM or STEM, used relatively low order diffraction spots in the Zero Order Laue Zone (ZOLZ). In other words, every study that did this is just based on the projection of the structure into a 2D-plane perpendicular to the beam direction. Whilst that has often revealed enough information for characterisation of the structures in question, it is also entirely possible that this information was incomplete because of the lack of access to the third dimension. Previously, one of the authors demonstrated the 3D connection of 60° domains in an antiferrolectric using shifts of high angle reflections in a very low angle first order Laue zone for the large unit cell in question [3], using a similar interpretation of the reciprocal lattice to Cai *et al.* [26]. However, in that study, no dark field imaging was performed using these shifted spots (nor could have been with the capabilities available at the time). More recently, Qiao *et al.* [27] performed a form of virtual dark field imaging from Position Averaged Convergent Beam Electron Diffraction (PACBED) data of reflections in a higher order Laue zone that appear at low temperatures due to Charged Density Waves giving a superlattice ordering. This is the only publication known to the authors of any form of Dark Field Imaging using specific reflections on a Higher Order Laue Zone (HOLZ) ring.

$La_2CoMnO_6$ (LCMO) is a ferromagnetic semiconductor with a range of interesting properties [28] and potential applications in spintronics [29] and resistive switching devices [30]. It can either crystallise in a disordered form with $Mn^{3+}$ and $Co^{3+}$ randomly distributed among the B-sites or an ordered form where the $Co^{2+}$ and $Mn^{4+}$ ions segregate to separate interpenetrating Face Centred sublattices, akin to the well-known rock-salt structure (although actually distorted into a monoclinic cell) [31]. This latter form is the more attractive one for magnetic properties. Egoavil *et al.* [32] showed that it was possible to grow partially ordered films on $SrTiO_3$ (STO) (111) via metalorganic aerosol deposition. Later, Wang *et al*. successfully grew it on $SrTiO_3$ (STO) (001) via chemical solution methods and achieved a high degree of order. However, STO has a larger lattice parameter than LCMO so places it in tensile in-plane strain in a film. To explore the alternative direction in strain, Lv *et al.*[33] showed theoretically that growth on $LaAlO_3$, which imposes compressive strain, changes the electronic structure and presumably properties quite markedly. Kleibeuker *et al.* showed that high quality thin films of the substance can be grown on $SrTiO_3$ (STO) or LSAT ($LaAlO_3$-$Sr_2TaAlO_6$) and that these have much better magnetic properties on (111) oriented substrates than the commonly-used (100) oriented substrates [34]. Moreover, the best properties were achieved on LSAT where there is compressive strain. This also seemed to coincide with the longest-range ordering of Co and Mn being seen on this substrate (as evidenced by atomic-resolution EELS mapping) [34]. However, the standard STEM techniques used at the time, HAADF imaging, ABF imaging, and EELS mapping, were not capable of revealing much crystallographic information about crystallographic domains or crystal orientations although it is clear from the low symmetry crystal structure that a number of possible orientations of the monoclinic cell to the substrate would be possible and therefore domains and domain boundaries should be present. It was speculated, based on strain arguments, that $[010]_M$ might prefer to lie in-plane on LSAT, but there was no evidence for this at the time. (In all cases, a subscript *M* is used to indicate indices with respect to the monoclinic cell described below).

It has long been recognised, however, that 3-dimensional information about structure and symmetry of materials is present in electron diffraction patterns at higher angles in HOLZ rings [35,36]. Moreover, it was realised some years ago that HOLZ rings can be used in STEM imaging to bring 3D crystallographic information into the images [37,38]. In recent years, this has been turned into a 4D-STEM imaging method using the intensity in a HOLZ ring to show the magnitudes of atom displacements associated with a specific crystal lattice ordering leading to a doubled periodicity *along the beam direction* [16,22,39,40]. Moreover, it was found that when a HOLZ ring arises due to back and forth atom movements along one specific direction, the intensity in the ring has a large azimuthal variation that can be fitted to reveal the direction of atom movements [41]. This approach was then applied to the LCMO film grown on LSAT by Kleibeuker *et al.* [34] and it was found that the 3D directions of the **b**-axes and magnitudes of the associated displacements La atoms could be mapped at atomic resolution [42]. In the small areas mapped, these displacements (which should be close to $[010]_M$) were in the interface plane, as predicted earlier by strain arguments [34].

In this work, we apply *Digital Dark Field* imaging to two thin films of LCMO grown on STO and LSAT using arrays of reflections in both the Zero and First Order Laue Zones (ZOLZ and FOLZ) together with examination of the azimuthal intensity distribution in the FOLZ. This reveals a much richer picture of domain structures and their connections to substrate-induced strain than was ever imagined previously, which gives a much clearer idea of why growth on LSAT is advantageous for ordering and magnetic properties.

## A. Possible orientations of LCMO cells on (111) perovskite substrates

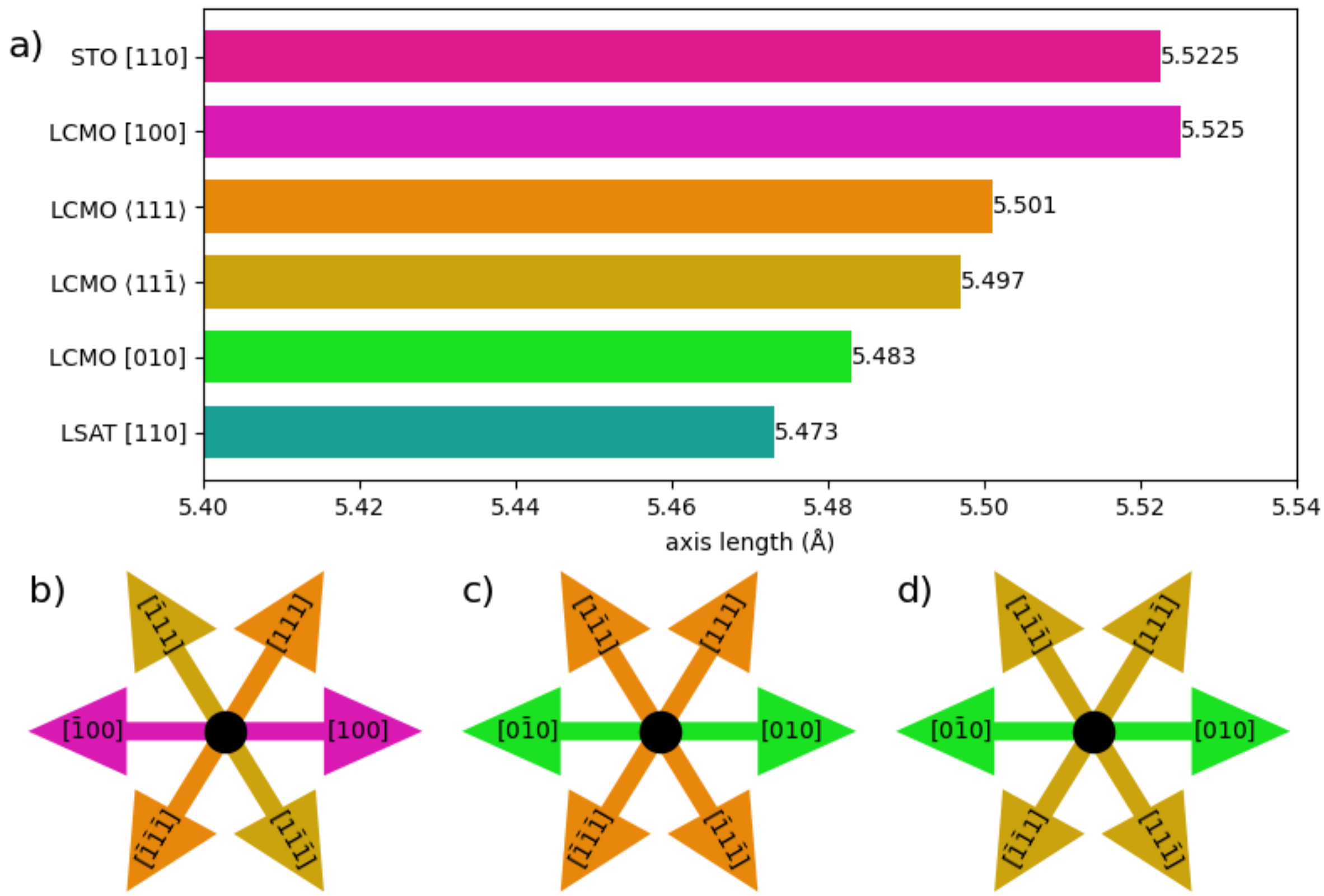

Figure 1: a) Lengths of different monoclinic axes equivalent to cubic perovskite [111] compared with the substrate vectors, and b)-d) Schematics of some different possible arrangements of these vectors of the monoclinic LCMO structure on a (111) surface of cubic perovskite, the colours of the arrows correlates to the colour scale in part a). Whilst other specific sets of indices are possible, these are the only three symmetrically distinct possibilities: b) **a**-axis in-plane ($(01\bar{1})_M$ growth plane); c) and d) **b**-axis in-plane ($(10\bar{1})_M$ and $(\bar{1}0\bar{1})_M$ growth planes).

LCMO has a monoclinic crystal structure based on a distortion of the $NdGaO_3$ orthorhombic perovskite structure, ideally with ordering of the Co and Mn sublattices in a rock-salt-like arrangement of two interpenetrating face-centred lattices. The lattice parameters are *a* = 5.52464Å, *b* = 5.48302Å, *c* = 7.7717Å, $\beta$ = 89.898° [31]. A key feature of this (and every structure based on $NdGaO_3$) is that the La (or other A-site) atoms move back and forth along the **b**-direction, something that we have already determined the direction of movement of in our previous publication [41] and which was recently mapped at atomic resolution [42,43]. To avoid confusion between references to directions and planes in cubic and monoclinic systems, cubic indices will have *C* subscripted, and monoclinic indices will have a *M* subscripted.

A $(111)_C$ oriented cubic substrate will have six $\langle 110\rangle_C$ directions in-plane, arranged at 60° intervals. There will be six more $\langle 110\rangle_C$ directions out of the plane, and each will be at 60° to two of the in-plane directions and at 90° to two more of them (which must lie antiparallel in-plane). These will align with equivalent directions to these in the monoclinic LCMO. The equivalent directions are:

$$[100]_M, [010]_M, [111]_M, [1\bar{1}1]_M, [11\bar{1}]_M, [1\bar{1}\bar{1}]_M$$

together with the negatives of these directions. It should be noted that in the *P2₁/n* structure of LCMO, the $[100]_M$ and $[010]_M$ directions are necessarily perpendicular, and therefore, one and only one of these two directions will lie in-plane. Schematics of vector orientations in the three crystallographically-distinct unit cell orientations possible on this surface are shown in Figure 1b)-d), one for $[100]_M$ in-plane, and two for $[010]_M$ in-plane. All of these diagrams use the colour scaling on Figure 1a), which shows the relative lengths of the different vectors (and a comparison to substrate vector lengths). Figure 1b) is equivalent to an $(011)_M$ interface plane, following the description of Kjaernes *et al.* [44] by growth plane in Miller indices of film structure (although modified for monoclinic, rather than orthorhombic), and is the only possibility for $[100]_m$ in-plane. Note that $[111]_M/[1\bar{1}1]_M$ and $[11\bar{1}]_M/[1\bar{1}\bar{1}]_M$ have different lengths because LCMO is monoclinic so there is a real difference between the arrangements in Figure 1 c) and d) (even if the strain difference is rather small). These could then be referred to as $(10\bar{1})_M$ and $(101)_M$ plane films following the description of Kjaernes *et al.* [44]. These different possibilities will be crucial in interpreting the structures revealed in the Results below.

## II. Methods

Two samples were grown on (111) substrates of STO and LSAT by pulsed laser deposition from prepared targets of $La_2CoMnO_6$. As detailed above in Figure 1, STO tends to put the film into tensile strain in the film plane, although placing the **a**-axis in-plane is low strain and slightly compressed (but other in-plane directions will still be in tensile strain). For LSAT, the film is in compressive strain along all in-plane directions. In both cases, the films were previously found to be fully coherent with the substrate with atomically sharp interfaces according to HAADF imaging. Long range Co-Mn double perovskite ordering was only found for the film on LSAT by atomic-resolution EELS mapping. Full details of growth, properties and atomic-resolution STEM characterisation are given in our previous publication [34]. Samples for electron microscopy were prepared by standard FIB liftout protocols.

Scanning transmission electron microscopy was performed using a JEOL ARM200F operated at 200 kV in TEM mode using the smallest spot size (5) and the smallest condenser aperture (10 μm) to produce the lowest convergence angle beam available of about 1.2 mrad semiconvergence angle, giving a minimum possible beam diameter ofof 2.1 nm. Scanning precession electron diffraction (SPED) datasets were acquired over suitable areas of film and substrate using a NanoMEGAS TopSpin system using a Quantum Detectors MerlinEM system for the diffraction pattern acquisition with a single 256×256 pixel Medipix-3 detector. For diffraction patterns containing higher order Laue zones, a short camera length of 30 cm was chosen to ensure the full angular range was seen on the detector, and a very small precession angle of 0.1° was selected to allow some of the benefits of precession in getting more uniform diffraction disks, whilst not mixing the different Laue zones (which happens at higher precession angles). In one case, a larger camera length of 80 cm was used to see more detail of the ZOLZ with a precession angle of 0.5°. All SPED data was exported as *.app5* files and converted to .hdf5 using Gary Paterson's *fpd* software [22].

The main 4D-STEM data processing was performed with *py4DSTEM* [17] (v. 0.4.18), including determination of list of diffraction peak positions (aka *pointslists*), determination of pattern centres, determination of lattice vectors for the perovskite lattice using the *strain* functions, and the Digital Dark Field imaging. All experimental diffraction patterns were plotted after rotation (with *scipy.ndimage*) to compensate for the calibrated rotation angle between the scanned image and diffraction (a calibration formula was previously determined using a MAG*I*CAL specimen [45] between a rotated scan of arbitrary scan rotation and the diffraction pattern orientation). Thus, any feature in any experimental diffraction pattern could be directly related to the spatial orientations of features in its partner image. Determination of the azimuthal modulation in the FOLZ from a list of diffraction spots was performed using an adapted version of the fitting previously published by Silinga *et al.* [41] written in python and using *scipy.optimize.curve_fit*. As shown in our recent work [42,43], the azimuthal variation comes from the antiparallel modulation of La atoms along the **b**-direction, at least for perfect crystal in the refined bulk structure. This atomic displacement is called **d**, which should be 0.0216**b** in the bulk, but because it could vary a little locally from bulk values in a thin film under the influence of substrate strain, crystallographic domain boundaries, crystal defects and surfaces, we refer to **d** in experimental determinations. A full code listing is provided in the data deposit accompanying this paper [46]. The reconstruction and visualisation of the 3D orientations of the **d**-axes was performed with standard python/numpy/matplotlib functions.

Dynamical simulation of diffraction patterns was performed using *py4DSTEM*, as above, using a Bloch wave method with the Weickenmeier-Kohl paramerisation of the elastic form factors [47], including the Debye-Waller factors, core absorption [48] and phonon scattering as an absorptive potential [47], as summarised by De Graef [49]. All simulations assume a basically parallel beam (no convergence is included) and sample the whole cell, and use <u> parameters calculated from the U parameters given by Bull *et al.* [31] of La: 0.063 Å, Co and Mn: 0.071 Å, O: 0.077 Å. A precession angle of 0.1° is used in the calculations, as in the experiments. A thickness of 50 nm was used for Figure 2 in the main text, although the effects of changing thickness are explored in the Supplemental Material. The form of the result does not change much with turning precession on or off, or with any other small changes to parameters. The aim is not for full matching of the whole pattern (including the diffuse scattering), but for showing the main trends of intensity in the FOLZ, which are not strongly affected by anything other than thickness. The spot radius in the plots depends on Intensity$^{0.5}$ to flatten the contrast a little and allow the FOLZ to be easily seen without oversaturating the ZOLZ. Full code is provided in the data deposit.

# III. Results and Discussion

## A. Diffraction pattern simulations to aid interpretation

Figure 2 shows simulations of 6 diffraction patterns from directions equivalent to $\langle 110 \rangle_{\text{cubic}}$ (technically, there are another 6 equivalent directions, which are the negatives of these, but these six simulations show the essential features). Two are for the principal lattice directions $[100]_M$ and $[010]_M$ (Figure 1a) and 1b)), and show no lower angle First Order Laue zone, but only a higher angle one at about 3.8-3.9 Å$^{-1}$. However, the other four (Figures 1c)-f)) have low angle First Order Laue zones at about 2.7 Å$^{-1}$. The symmetry reduction and increase in cell size means that none of these latter four has twofold symmetry or any true mirrors, once we consider the full pattern including the Higher Order Laue Zones [36]. The shift of pattern origin for the *Nth* order Laue zone in this structure with $a \approx b \approx \sqrt{2}a_{\text{cubic}}, c \approx 2a_{\text{cubic}}$ is approximately:

$$\mathbf{g}_B = h_B k_B l_B = \frac{Nu}{4}\frac{Nv}{4}\frac{Nw}{2} \quad \text{for} \quad \mathbf{B} = [uvw] \qquad [1]$$

where $\mathbf{g}_B$ is a direction in reciprocal space, expressed in Miller indices, parallel to the beam direction, $\mathbf{B}$, expressed as a normal real-space vector. $\mathbf{g}_B$ is given with a magnitude equal to the spacing between the 0$^{\text{th}}$ and N$^{\text{th}}$ reciprocal lattice layers, and the relationship between the indices of parallel real and reciprocal space vectors is the normal one based on the standard definitions of **a**$^*$, **b**$^*$ and **c**$^*$ in terms of **a**, **b** and **c**, the basis vectors for the unit cell in real space. So, for example, [111] gives $\mathbf{g}_B = \frac{1}{4}\frac{1}{4}\frac{1}{2}$ for the centre of the FOLZ. This means that ¼ shifts are going to be found along $\langle 112 \rangle$ directions in the FOLZ, but these could be either up or down in the plane of the figure, depending on the rotational orientation of the diffraction pattern. Figures 1c)-f) show different choices for the $\langle 112 \rangle$ direction, always pointing downwards in the figure. For 1d) and 1f), the ¼ shift follows this $\langle 112 \rangle$ direction, whereas in 1c) and 1e) it is antiparallel, which can be equivalently expressed as -¼ or ¾ (the latter choice is made in this work). This shift of spots in the FOLZ with respect to the ZOLZ is clearly linked to the direction of the crystallographic **b**-axis. In each case, the direction in which $[010]$ or $[0\bar{1}0]$ is pointing out of the page towards the reader is shown with an orange arrow (the beam direction out of the page towards the reader is given in every subfigure title). In all cases, if the ¼ shift downwards is present, **b** is pointing up somewhere in the upper half of the pattern. If the ¾ shift is present, **b** is pointing up somewhere in the lower half of the pattern.

To compare with our recent publications where the intensity modulation in the FOLZ was fitted [41] and then used to determine the absolute direction of **b** at atomic resolution [42,43], the brightest quadrant was determined and indicated in pink. For the thickness used in the simulation, this was always the same quadrant as contained the **b** axis pointing up out of the page. However, further investigation as shown in Figure S1 of the Supplemental Material shows that the contrast modulation is thickness dependent – occasionally, at some thicknesses, the brightest quadrant is the opposite one to that containing the **b** axis pointing up out of the page. Thus, whilst mapping the azimuthal modulation determines the in-plane direction with a 180° ambiguity, determining the absolute sign of the 3D direction is most robust in nanobeam diffraction by combining the directional peak of modulation (which narrows it down to two quadrants) and the direction of the ¼ shift (which then further narrows it down to one). This, therefore, gives a method for determining the 3D orientation of the **b**-axis from nanobeam diffraction patterns. Firstly, we can map the direction of the shift using Digital Dark Field Imaging [25]. Secondly, we can map the peak modulation direction using the azimuthal fitting algorithm used previously [41], albeit adapted for datasets that are discrete lists of diffraction spot positions and intensities from disc detection applied to nanodiffraction [25] (aka *pointslists* [17,24]) rather than a continuous intensity distribution to get the peak intensity direction.

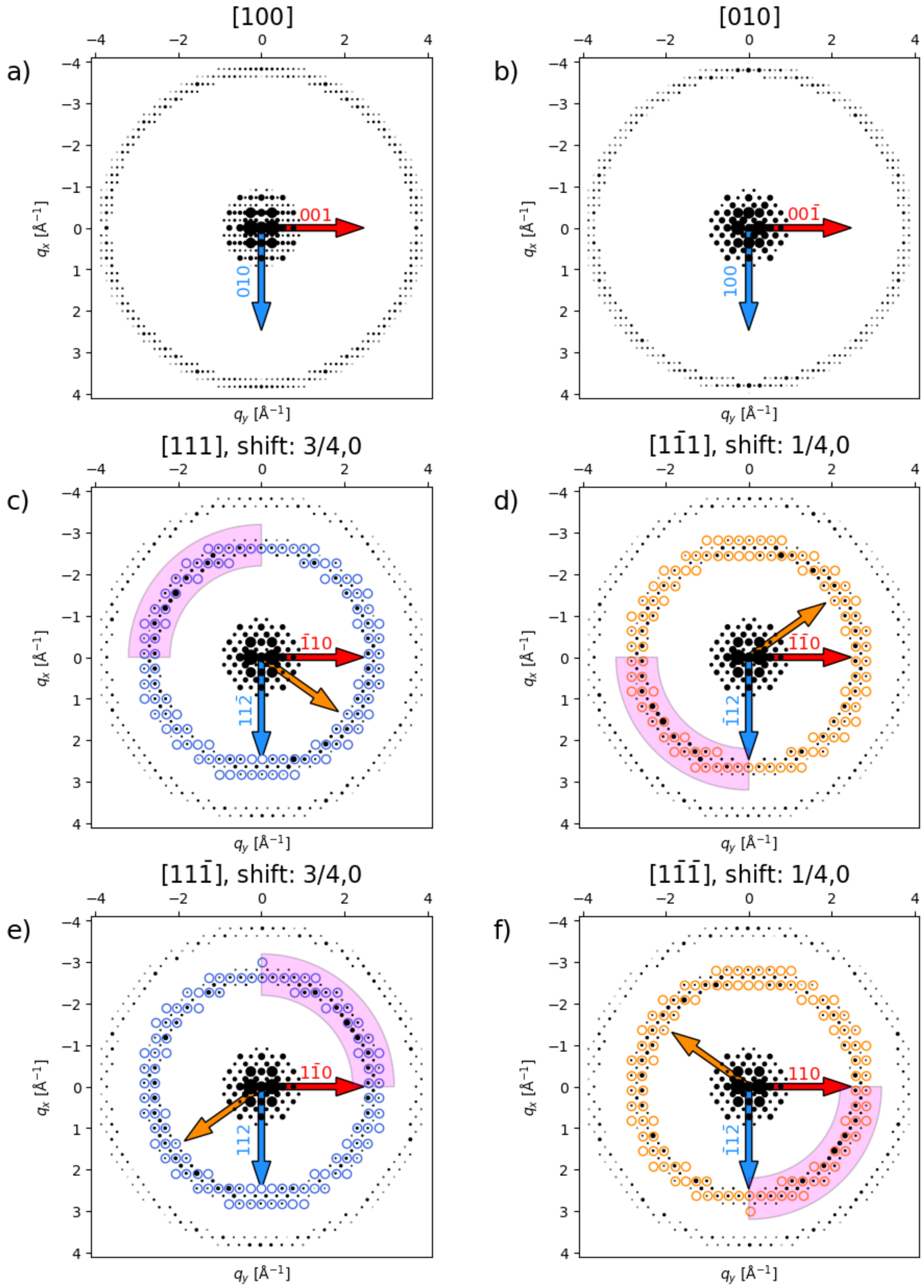

Figure 2: Dynamical simulations of diffraction patterns for six distinct orthorhombic directions equivalent to $\langle 110 \rangle_{\mathrm{cubic}}$ for a crystal thickness of 50nm. In each case the directions of $\mathbf{g}_1$ (vertical, blue) and $\mathbf{g}_2$ vectors (horizontal, red) are arrowed with indices (although absolute length is magnified to make the arrows easy to see). Five of the six

have similar ZOLZ patterns; the latter four have an inner FOLZ. The shift of the rectangular lattice of the ZOLZ to the pattern is shown, either a ¼ shift downwards (and FOLZ reflections circled in orange) or a ¾ shift downwards (equivalent to a ¼ shift upwards, and FOLZ reflections circled in blue). The direction in which the **b**-axis of the structure points upwards out of the page is shown with an orange arrow. The brightest quadrant of the FOLZ is highlighted in pink.

The angles between $\mathbf{g}_1$ and $\mathbf{g}_2$ are given in Table 1 for each of the diffraction patterns of Figure 2, and this information is useful in later discussion and interpretation of experimental results.

| **B** | $\mathbf{g}_1$ | $\mathbf{g}_2$ | Angle (°) |
|---|---|---|---|
| $[100]$ | $020$ | $002$ | 90.00 |
| $[010]$ | $200$ | $00\bar{2}$ | 89.90 |
| $[111]$ | $11\bar{2}$ | $\bar{1}10$ | 89.74 |
| $[1\bar{1}1]$ | $\bar{1}12$ | $\bar{1}\bar{1}0$ | 90.26 |
| $[11\bar{1}]$ | $112$ | $1\bar{1}0$ | 90.36 |
| $[1\bar{1}\bar{1}]$ | $\bar{1}1\bar{2}$ | $110$ | 89.64 |

Table 1: Angles between the basis vectors for each of the diffraction patterns of Figure 2.

## B. Digital Dark Field Imaging of the Film grown on LSAT

Figure 3 shows Digital Dark field (DDF) images and corresponding diffraction patterns from a few areas in a thin film grown on LSAT. Figure 3a) shows an image formed with just a rectangular array of primitive perovskite reflections in the ZOLZ; and the corresponding diffraction pattern for the image position indicated with a white spot is shown in Figure 3g), with red-orange ring overlays for each diffraction spot position included in the summation. Another diffraction pattern marked in the same way is shown in Figure 3h), from the point marked with the white spot in Figure 3d). This image of Figure 3a) should therefore highlight all perovskite areas with the same crystal orientation. Most areas of both film and substrate are bright in this image, as one would expect for an epitaxial perovskite film on a perovskite substrate. The top of the image area is dark as this is either carbon or deposited platinum and is not imaged by this array of dark field apertures. There is a dark area to the lower right, which is probably a CoO inclusion, as noted previously [34]. There is also a general intensity gradient from brighter on the left to less bright on the right, which is simply from a thickness gradient in the specimen (which causes more intensity to be scattered outside of coherent diffraction spots).

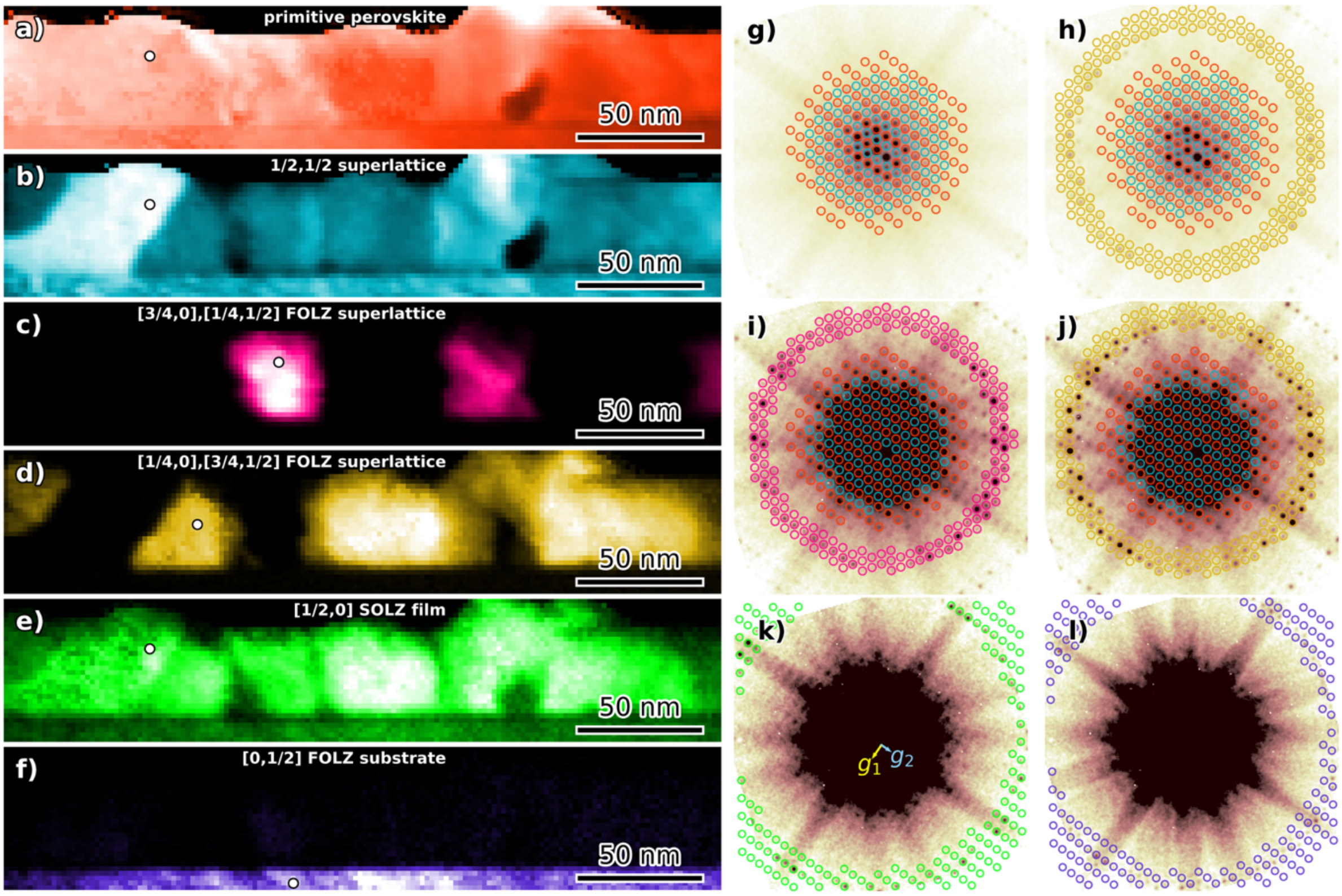

Figure 3: Digital Dark Field (DDF) Images from the LCMO film grown on LSAT for: a) primitive perovskite reflections; b) reflections shifted by ½,½ in the ZOLZ; c) reflections shifted by ¾,0 and ¼,½ in the FOLZ; d) reflections shifted by ¼,0 and ¾,½ in the FOLZ; e) reflections shifted by ½,0 in the outer Laue zone (SOLZ for $\langle 111\rangle$ or $\langle 11\bar{1}\rangle$); f) reflections shifted by 0,½ in the outer Laue zone (FOLZ for substrate). Corresponding diffraction patterns with aperture overlays in suitable colours to match the images: g)-h) diffraction patterns from the white point in a) with overlays for primitive perovskite (red-orange) and ½,½ spots (turquoise) and the white point in d) with the addition of a gold overlay for the FOLZ, in both cases saturation set to allow the ZOLZ to be visible; i)-j) diffraction pattern from the marked points in c) and d) with FOLZ overlays in hot pink and gold, saturated more to allow the FOLZ to be more clearly seen; k)-l) diffraction patterns from the marked points in e) and f) saturated even more strongly to bring out the detail in the outer HOLZ ring, together with overviews in green and purple-blue to match ½,0 and 0,½ shifts

Figure 3b) shows an image formed with a rectangular array of "apertures" in the ZOLZ with a ½,½ shift of the origin to pick up the superlattice spots expected for the orientations shown in Figure 2b-f), and two corresponding diffraction patterns are shown in Figures 3g) and 3h) with turquoise overlays indicating the aperture positions (extracted from the white spot positions in 3a) and 3d)). The $\mathbf{g}_1$ and $\mathbf{g}_2$ vectors are drawn on Figure 3k) (to scale) to show the directions from which these shifts are calculated. Again, almost all areas of the film and substrate appear bright with this set of reflections, indicating they are present for any and all domains of both substrate and film in this sample orientation. Again, the only obvious exception is the CoO inclusion to the lower right. However, one area of the film to the left-hand side is much brighter than everywhere else in the film. This is the area from which the diffraction pattern of Figure 3a) was extracted, and it can be clearly seen that there is no inner FOLZ and a fairly bright set of ½,½-shifted reflections in the ZOLZ. This corresponds with expectations for [010] as shown in Figure 2b). The fact that Figure 3b) is bright in patches of the substrate indicates that this is not a primitive perovskite for the most part, but has some additional unit cell ordering leading to the appearance of this superlattice spot – this would certainly be expected in $LaAlO_3$, which LSAT is based upon.

However, most areas of the film do contain an additional inner FOLZ ring. Figure 3c) is made for FOLZ reflections with a shift of ¾,0 and ¼,½ (these correspond to the primitive array of spots and the ½,½-shifted array of superlattice spots in the ZOLZ, all with a ¾,0 shift applied for the FOLZ). (All shifts refer to fractions of the $\mathbf{g}_1$ and $\mathbf{g}_2$ drawn in Figure 3k)). The location from which the exemplar diffraction pattern of Figure 3i) in extracted is shown with a white dot in Figure 3c), with the array of aperture positions shown in hot pink. Clearly there is a strong FOLZ ring in the diffraction pattern and two particular areas of the film are seen in the image that are bright, which must be domains with a specific orientation related to one of Figures 2c)-f). The substrate is dark in Figure 3c).

Similarly, Figure 3d) is made for FOLZ reflections with a shift of ¼,0 (corresponding to the primitive array in the ZOLZ) and ¾,½ (corresponding to the superlattice of ½,½-shifted array in the ZOLZ). The location from which the exemplar diffraction pattern of Figure 3j) was extracted is shown with a white dot. The array of aperture positions is shown in gold on the diffraction pattern. Again, the FOLZ ring is strong and several bright areas of film the film are seen in the image, corresponding to specific domains. And as for Figure 3c), the substrate is dark. So, this inner FOLZ is only present in the film.

It should also be noted that the brightest parts of the FOLZ ring for both Figures 3c) and 3d) are to the left and right of the rotated diffraction patterns, indicating a **b**-axis in the plane of the specimen, which is also consistent with the single $[010]$ domain seen in Figure 3b). The frequent reversal between ¾,0 and ¼,0 orderings suggests that the orientation of this **b**-axis rotates frequently in the plane of the film between successive domains, but a full 3D rendering will be calculated below.

Figure 3e) was made with a set of apertures corresponding to the SOLZ (½,0 shift) for the $\langle 111 \rangle_M$ or $\langle 11\bar{1} \rangle_M$ directions (equivalent to the FOLZ for the primitive perovskite structure or for the $[100]_M$ or $[010]_M$ directions), and a corresponding diffraction pattern is shown in Figure 3k). All film domains light up in this image, and the diffraction pattern is shown with green-blue overlays for the apertures. The substrate was darker in this image and the FOLZ is weakened here – which is probably a result of significant local compositional disorder in the LSAT. A similar effect was seen previously for La(Sr,Mn)$O_3$ [39].

Nevertheless, additional features were seen in the FOLZ for the substrate, the diffraction pattern being shown in Figure 3l). The additional FOLZ reflections are identified by the purple-blue apertures in an array corresponding to 0,½ shifts. Forming a DDF image from this results in something that only lights up the substrate. Thus, it seems that there is an ordering in the substrate structure which is only identified in the FOLZ, and which is different to that seen in any of the film structures along any of the directions imaged. It is likely this is related to the $LaAlO_3$ structure that the LSAT structure is based upon.

From a technical point of view on the imaging, it is also interesting to note how well the DDF technique works over such a large range of different intensity levels in the spots used. Figure 3a) is about 250 times brighter than 3f) in absolute intensity and yet the DDF imaging continues to work well with acceptable Signal to Noise and Signal to Background Ratios. See Figure S3 in the Supplemental Material for further details.

## C. Digital Dark Field Imaging of the Film grown on STO

Figure 4 shows DDF images and their corresponding diffraction patterns for the LCMO film grown on STO, and the same figure parts are calculated with the same arrays of apertures for Figures 4a)-d) to those used for Figures 3a)-d). Figure 4a) is a DDF image for an array of spots in the ZOLZ corresponding to primitive perovskite and this lights up both film and substrate, as expected. This array of aperture positions is shown superimposed on diffraction patterns for two different film areas in Figs 4f) and g) using red-orange rimmed discs.

Figure 4b) shows a DDF image calculated for ZOLZ reflections for a ½,½ shifted array, which is superlattice spots only seen for the orientations simulated in Figure 2b)-f). The corresponding array of aperture positions in the diffraction pattern are shown in Figs 4f) and g) using turquoise-rimmed discs. As for Figure 3, the $\mathbf{g}_1$ and $\mathbf{g}_2$ vectors are drawn on Figure 4k) to scale to show their size and orientations, and all shifts calculated as multiple of these vectors. This ½,½ shifted array lights up most of the film (save a patch in the centre) but none of the substrate. The substrate not lighting up is no surprise as it should be simple cubic $SrTiO_3$ with no superlattice spots at all.

Figure 4c) shows a DDF image from the FOLZ for the ¾,0 shift. Unlike for the film on LSAT, the ¼,½ shifted array was not calculated as such spots seemed extremely weak or invisible. This is possibly a sign of a further subtle structural alteration from the bulk structure. The array of aperture positions is shown using hot pink overlays on the diffraction patterns of Figs 4f) and 4h) which correspond to the position shown with the white dot in Figure 4a-c). The FOLZ spots are hard to see in the diffraction patterns as shown in Figure 4f) with contrast set so that the ZOLZ is not oversaturated, emphasising that the FOLZ spots are very weak. Nevertheless, with an altered saturation point in Figure 4h), the FOLZ spots are clearly visible and match the positions of the ¾,0 – shifted array. Using this array of aperture positions to form a DDF image lights up the vast majority of the area lit up in Figure 4b).

Figure 4d) shows a DDF image from the FOLZ for the ¼,0 shift and, in this case, only a tiny triangular notch is lit up with this particular shift. Again, the second array to go with this, ¾,½, was not calculated as spots in this array are weak or invisible in the diffraction patterns. The array of aperture positions is overlaid on the diffraction patterns of Figure 4g) and 4i) in gold circles. As before, the FOLZ is hard to see at a contrast that allows good visibility of the ZOLZ, but by saturating the ZOLZ, the FOLZ stands out in Figure 4i), and the array matches well.

Whilst superlattice spots in the ZOLZ with 0,½ shifts could be discerned by eye in this diffraction dataset, they were so close to the much brighter primitive perovskite spots that automated peak detection was not reliable. Thus, a dataset at a longer camera length used in our previous work [25] was used to make this image, and after a little cropping and correction of drift distortion in that dataset, this is shown in Figure 4e). The diffraction pattern with apple-green overlays is shown in Figure 4j). This clearly demonstrates that the central domain of the film is $[100]$ oriented. A high angle diffraction pattern for the same point in the main dataset is shown in Figure 4k) and shows that there is no low-angle FOLZ here, as would be expected from the simulation of Figure 2a).

The outer Laue Zone (SOLZ for $\langle 111 \rangle$ or $\langle 11\bar{1} \rangle$) was generally weak in this dataset so was not used for making a DDF image, unlike in Figure 3e).

As for the azimuthal angle at which the intensity peaks in the FOLZ, this appears to be somewhere round about 100-110° and 280-290° ACW from horizontal right (which is the consistent definition in this paper for 0°) for the ¾,0 shift area in Figure 4c), suggesting a **b**-axis orientation not in the film plane, which means the **a**-axis must lie in-plane. One $[100]$ domain is also seen which agrees with this conclusion of the **a**-axis being preferentially in plane for the film grown on $SrTiO_3$. However, for the triangular domain of Figure 4d), the peak intensity lies around 0° / 180°, suggesting **b**-axis in-plane.

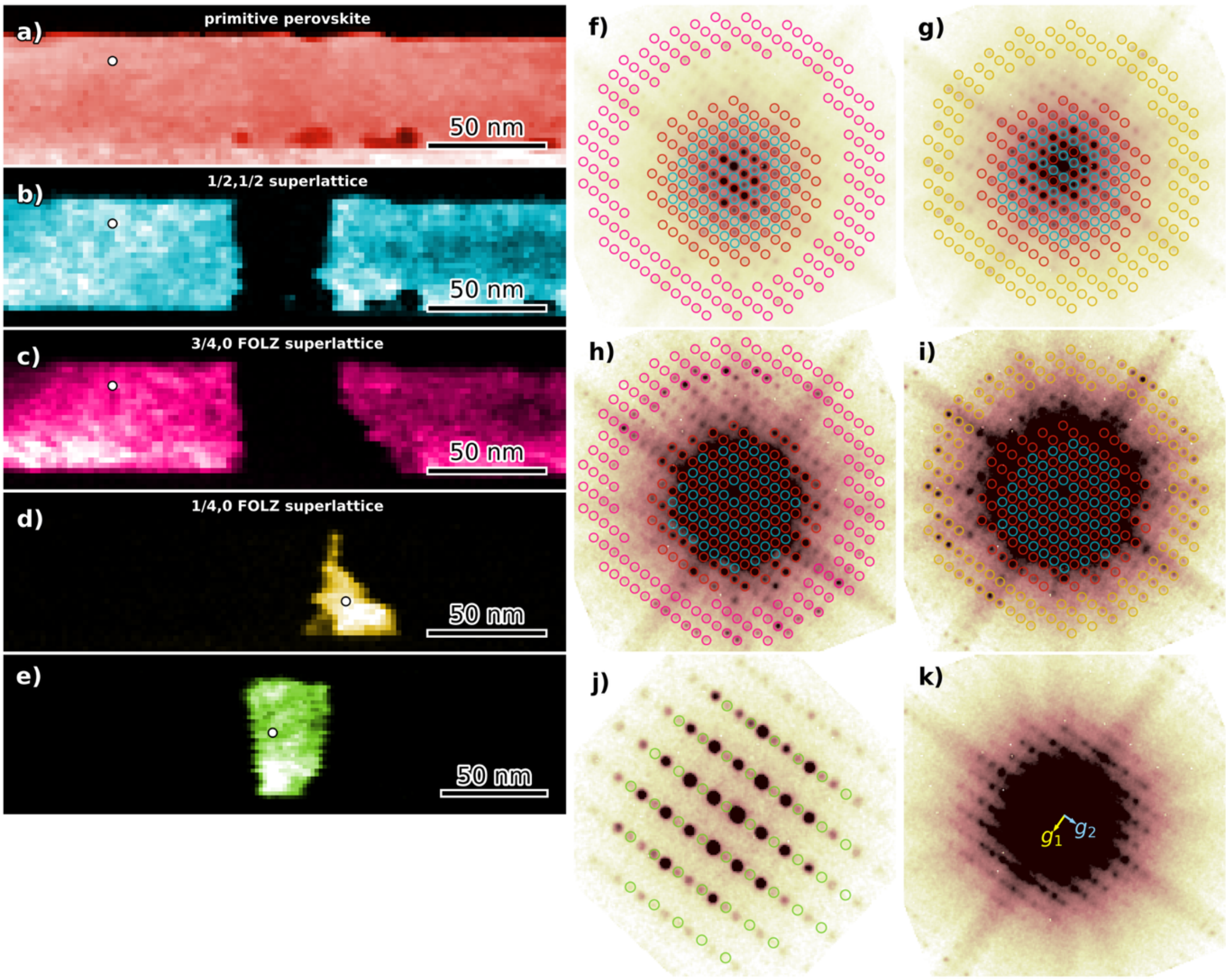

Figure 4: Digital Dark Field (DDF) Images and corresponding exemplar diffraction patterns with overlays for the DDF apertures from the LCMO film grown on STO. DDF images for: a) primitive perovskite reflections; b) reflections shifted by ½,½ in the ZOLZ; c) reflections shifted by ¾,0 in the FOLZ; d) reflections shifted by ¼,0 in the FOLZ; e) DDF image and diffraction pattern from a different dataset at a different camera length on this area for reflections shifted by 0,½ in the ZOLZ. Diffraction patterns from: f), h) the pixel indicated by the white dot in a-c) with contrast altered to allow either f) the ZOLZ or h) the FOLZ to be clearly seen; the same aperture overlays are shown for the spots seen here on both versions with a (3/4,0) shifted array in the FOLZ; g), i) the pixel indicated by the white dot in d), again at lower saturation g) to allow the ZOLZ to be seen and higher saturation to allow the FOLZ to be clearly seen i), the same overlays fit in the ZOLZ but the FOLZ overlay is the (1/4,0) shifted one; j) diffraction pattern from the white dot in e) from a separate dataset, with (0,1/2) shifted array of overlays superimposed, k) diffraction pattern from the same location in the HOLZ dataset showing no low-angle FOLZ ring. All high angle diffraction patterns are displayed with the displayed intensity given by $counts^{0.1}$ to flatten contrast. j) is displayed with intensity = $counts^{0.33}$.

## D. 3D crystal orientations – reconstruction, reasons and consequences

Combining the information encoded in the DDF images from the FOLZ showing the direction of the ¼ shift, and the information from fitting the azimuthal intensity variation, it is possible to construct 3D colour maps of **d**-axis orientation. More details about all the logical steps are provided in the Data Deposit which includes the full calculation [46]. However, the principles are summarised here. Areas where there is a ¾,0 shift (when $\mathbf{g}_1$ points down-left) must have the upwards inclination of the **d**-direction either oriented roughly leftwards or downwards (following Figure 1c) and 1e)). Areas where there is a ¼,0 shift must have the upwards inclination of the **d**-direction either oriented roughly rightwards or upwards (following Figure 1c) and 1e)). So, the relevant images (Figs 3c) and 4c) for the ¾,0 shift, and Figs 3d) and 4d) for the ¼,0 shift) can be thresholded to find areas where a specific Laue zone ordering is strong. The diffraction spots in the FOLZ can then be extracted and the fit previously used for the azimuthal intensity variation

$$I(\phi) = A_2 \cos^2(\phi - \phi_2) + A_1 \cos(\phi - \phi_1) + B \quad [2]$$

can then be applied to the data, fitting intensity, $I(\phi)$, of spots against their azimuthal angle, $\phi$ [41-43]. $A_2$ and $\phi_2$ are the amplitude and angular offset of the two-fold modulation (i.e. peaking twice, 180° apart), $A_1$ and $\phi_1$ are the amplitude and angular offset of the one-fold modulation (i.e. peaking once in 360°) and $B$ is the offset from zero (this is a change in parameter names from Silinga *et al.* [41]). The $\phi_2$ is then used to determine the **d**-axis orientation with an uncertainty of 180°, and this ambiguity is resolved depending on whether it is in a ¾,0 (left, down) or ¼,0 (right, up) domain. Where a $[010]$ domain is found, it is clear that **b**-axis orientation is vertical, so we assume **d** is vertical. Where a $[100]$ domain is found, the **b**-axis lies along $\pm\mathbf{g}_1$ and this lies in-plane, so we assume **d** lies along this direction. So, using image processing, we can then superimpose the different domains into a single image with colours related to **d**-axis orientation using a suitable colour-wheel representation of **d**-axis orientations in the upper hemisphere of possible orientations (i.e. out of the page towards the reader). Arrows can also be added showing the in-plane component of the **d**-axis using some reasonable sampling of just a fraction of image positions.

Figure 5 shows this reconstruction for the film grown on LSAT, and the clear tendency is to form fairly equiaxed domains of the order of 50 nm in size, all with the **d**-axis close to in-plane, whether red (up-right), cyan (up-left) or white (parallel to beam direction). Nevertheless, there are some slight tilts away from being perfectly parallel to the interface, and this may suggest some slight local deviations of **d** from the **b**-direction.

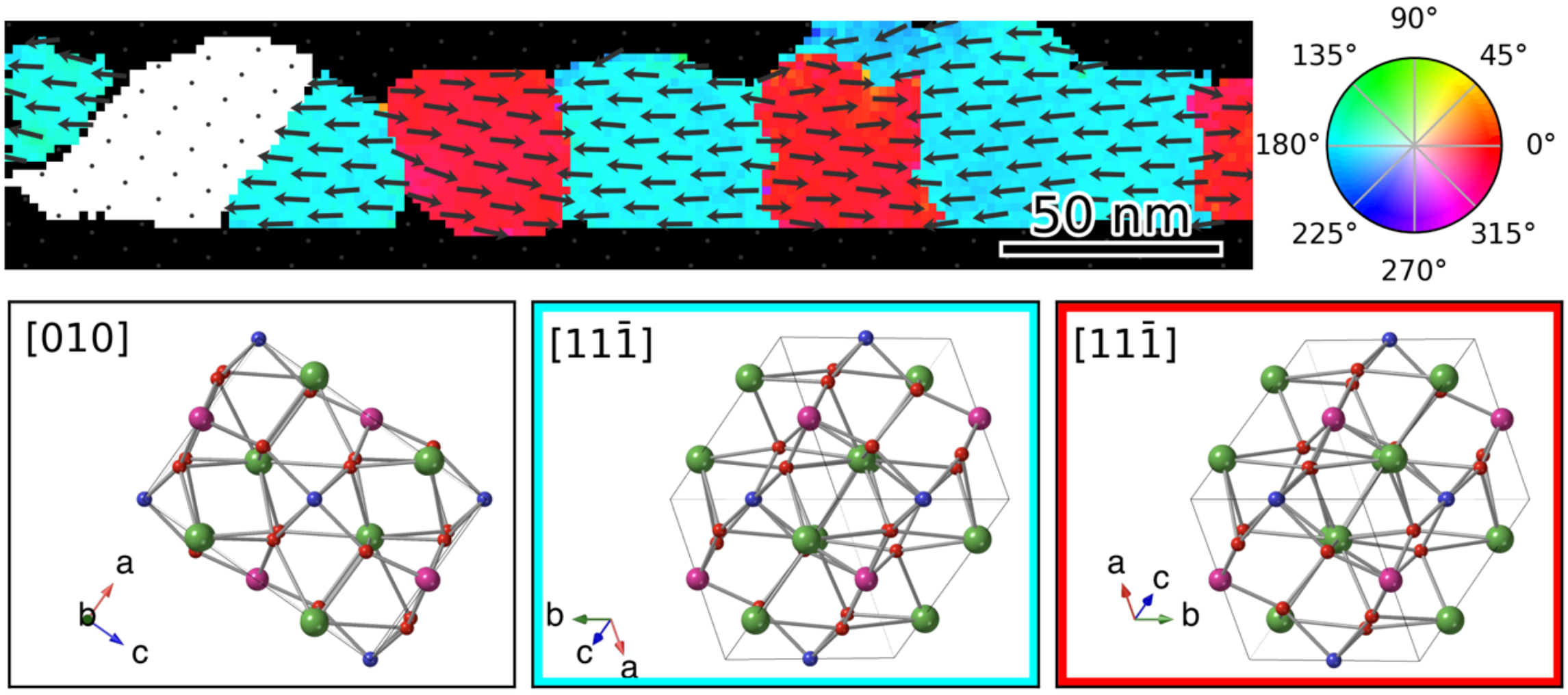

Figure 5: 3D orientations of **d**-directions in the LCMO film grown on LSAT. The colour scale indicates 3D orientation in the upper hemisphere (out of the page), and the arrows are superimposed to make this even more unambiguous. Crystal models for the three orientations are shown below, with colour coded frames matching the domain colours.

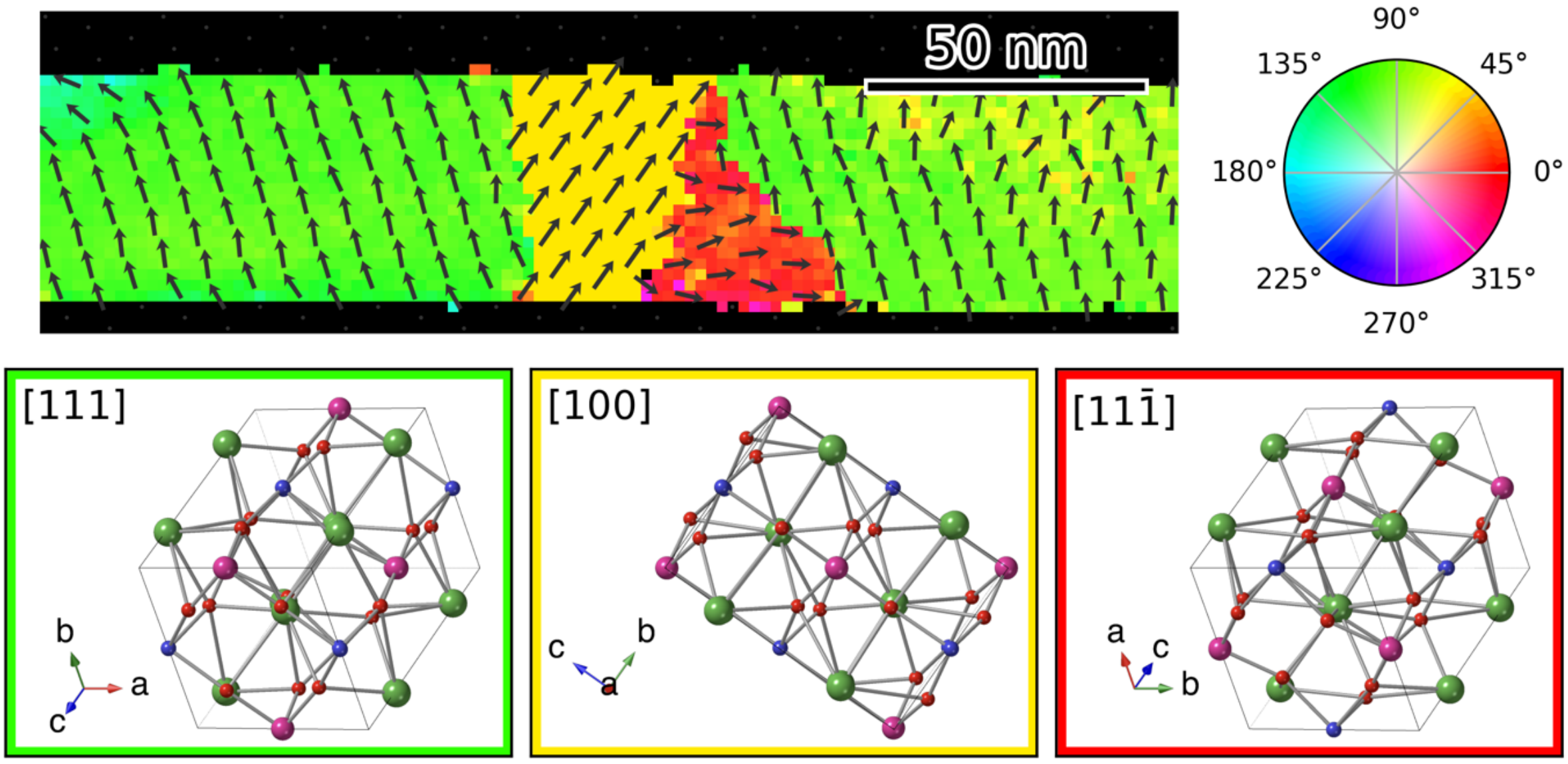

Figure 6: 3D orientations of **d**-directions in the LCMO film grown on STO. The colour scale indicates 3D orientation in the upper hemisphere (out of the page), and the arrows are superimposed to make this even more unambiguous. Crystal models for the three orientations are shown below, with colour coded frames matching the domain colours.

Figure 6 shows the same reconstruction for the film grown on STO. In this case, the clear tendency of the **d**-directions, and therefore **b**-axes to line up in directions out of the film plane (technically, at ~109° to film plane) as seen in the green areas or ~55° in the yellow domain, although there is the one anomalous area where the **d**-axes have lined up close to parallel to the film plane in the red notch. Quite why no arrows pointing downwards and out of the page are seen is unclear, as it would seem that this

**b**-axis orientation should be equally likely if **b** tends to point out-of-plane. Perhaps it is the case that the domains are relatively large, and such domains may be elsewhere in the film. Certainly, the density of domain boundaries seems smaller than in the LSAT film for some reason. Maybe the strains caused by domain formation out-of-plane would lead to highly strained domain boundaries which could disfavour that formation.

Now the reason for this change in preferred crystallographic orientation of domains between the two substrates is fairly clear when one considers the differences in lattice parameters, as summarised in Figure 1a). For growth on LSAT, whichever of the possible equivalents to $[110]_C$ is in-plane, these directions would all be slightly long and the film will be in compressive strain in-plane, possibly with some compensatory out-of-plane expansion [50]. This is borne out in the strain maps shown in Figure 7 (the reference is the grey box in the substrate at the bottom of the scan), where the out-of-plane strain in Figure 7a) is strongly positive (around 1.0-1.5%). The lowest strain set of vectors in-plane from Figure 1 would be $\langle 010 \rangle$, $\langle 11\bar{1} \rangle$ and $\langle 11\bar{1} \rangle$ (Figure 1d)), corresponding to $\{101\}_M$ growth, closely followed by $\langle 010 \rangle$, $\langle 111 \rangle$ and $\langle 111 \rangle$ (Figure 1c)), corresponding to $\{10\bar{1}\}_M$ growth. This is exactly the preference seen in Figure 5.

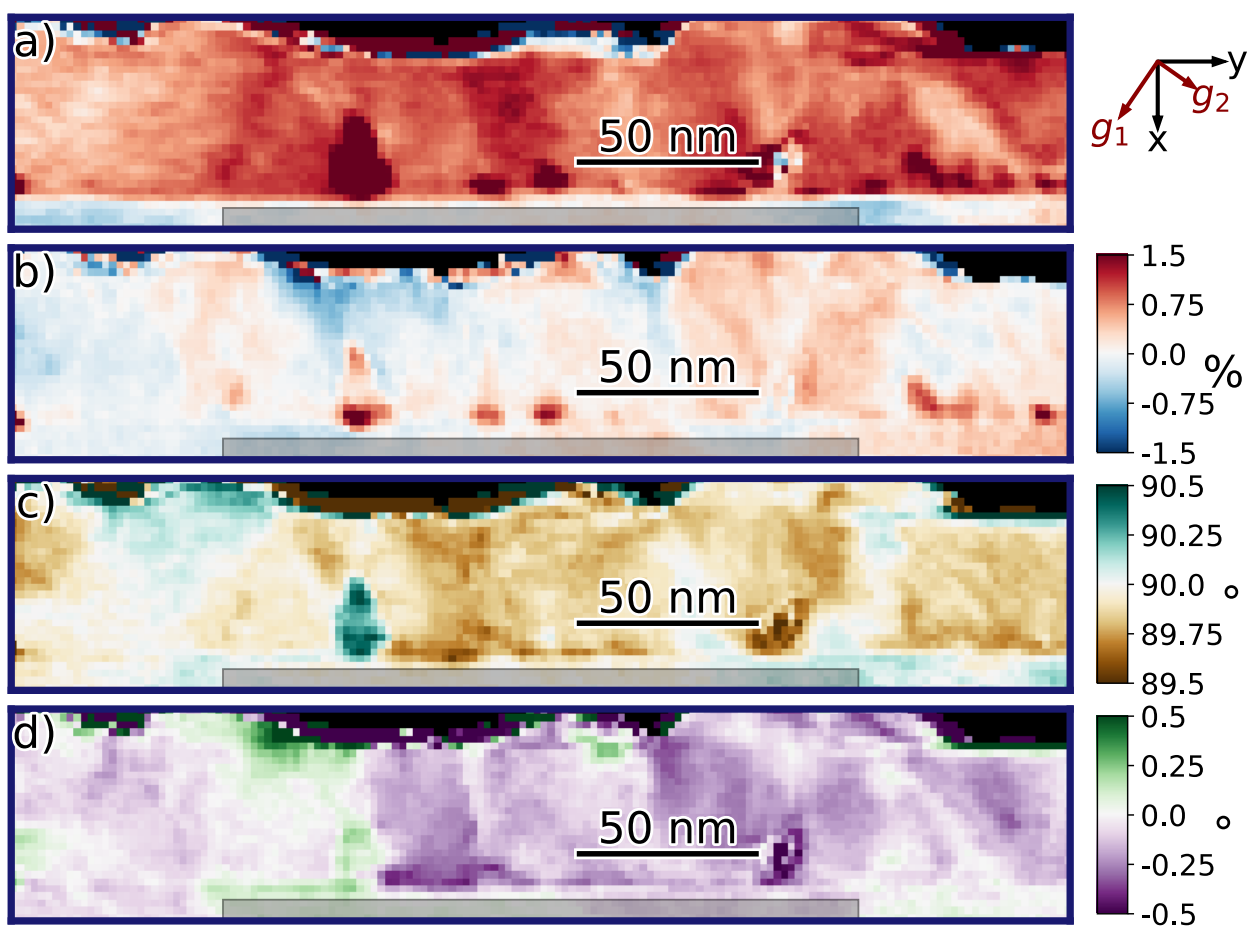

Figure 7: Strain maps for the LCMO film on LSAT. The **g**-vectors of the diffraction pattern and the axes for the strain calculation are shown. All quantities are calculated with reference to the grey box covering the substrate at the bottom: a) out-of-plane ($\varepsilon_{xx}$); b) in-plane ($\varepsilon_{yy}$) ; c) angle between $g$-vectors ($\phi$), calculated from shear ($\varepsilon_{xy}$) ; d) lattice rotation ($\theta$).

The out-of-plane spacing would be 4.499Å for $\{10\bar{1}\}_M$ and 4.507Å for $\{101\}_M$, if totally unstrained. The average elastic strains imposed in-plane on the film from the substrate (Figure 1c and 1d) would be $-0.48\%$ for $\{10\bar{1}\}_M$ and $-0.40\%$ for $\{101\}_M$. We assume a simple rectangular shape change (i.e. $\frac{dx}{x} = -\frac{dy}{y}$ or Poisson's ratio $\upsilon = 1$) because the sample is thin in the viewing direction and any strain in this direction can be released by free expansion of the surfaces. This gives total out-of-plane strains of $+1.21\%$ and $+1.63\%$, respectively. These seem the right order of magnitude, although the observed strains in the $\varepsilon_{xx}$ map seem closer to those for $\{10\bar{1}\}_M$ than for $\{101\}_M$. If $\upsilon < 1$, these calculated values would be higher.

There is a little variation in in-plane strain from one domain to another in Figure 7b) and it is close to zero, suggesting coherent growth in-plane, except for CoO at the interface, consistent with the atomic resolution imaging published previously [34].

By converting the shear strain to an angle between vectors and plotting this in Figure 7c), it is possible to compare the angles to those calculated in Table 1. Much of the film is therefore consistent with $\langle 111 \rangle$ or $\langle 1\bar{1}\bar{1} \rangle$ but distinguishing the two zone axes and the monoclinic distortion seems consistent with that in the bulk refinement [31]. In the area of the $\langle 010 \rangle$ domain on the left (white in Figure 5) the angle seems to be closer to 90°, as should be the case (89.9/90.1°). However, the angle between vectors in the strain maps seems consistently less than 90° throughout the majority of the film, which would be in accordance with positive strain out-of-plane. There is also a little lattice rotation from one domain to another in Figure 7d).

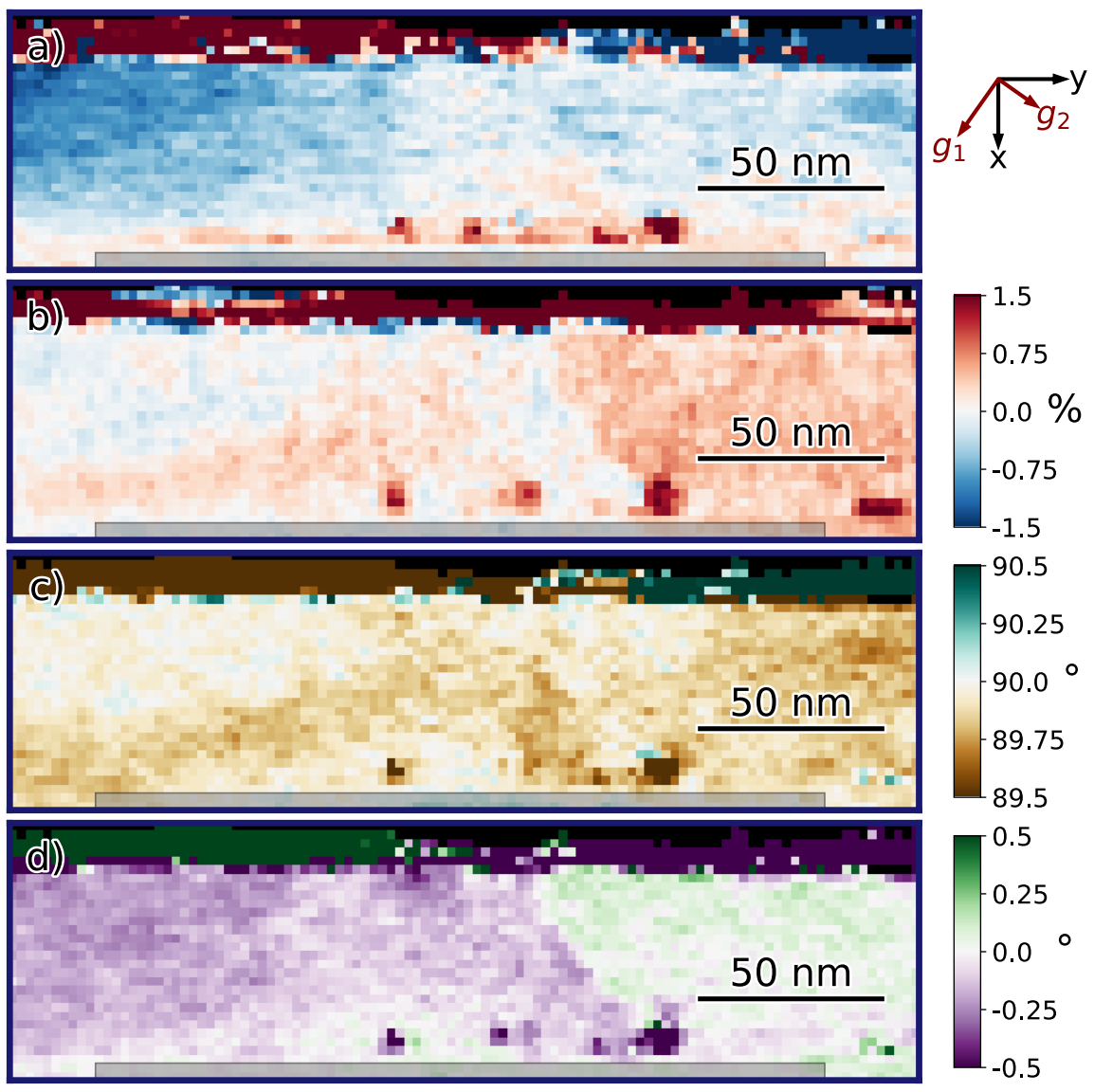

Figure 8: Strain maps for the LCMO film on STO. The **g**-vectors of the diffraction pattern and the axes for the strain calculation are shown. All quantities are calculated with reference to the grey box covering the substrate at the bottom: a) out-of-plane ($\varepsilon_{xx}$); b) in-plane ($\varepsilon_{yy}$) ; c) angle between $g$-vectors ($\phi$), calculated from shear ($\varepsilon_{xy}$) ; d) lattice rotation ($\theta$).

For growth on STO, most of the possible equivalents to $[110]_C$ for LCMO are slightly short, except for $\langle 100 \rangle_M$. So, the lowest strain option is $\langle 100 \rangle_M$, $\langle 111 \rangle_M$ and $\langle 11\bar{1} \rangle_M$ directions in-plane (Figure 1b)). This is the preference seen in Figure 6, although one minority area has nucleated a $\langle 010 \rangle_M$ in-plane. It is unclear why this is the case, but maybe strain from surrounding domains stabilised an unusual strain state at this junction.

These conclusions are supported by the strain map shown in Figure 8 (the reference is the substrate at the bottom of the scan). There is little difference in in-plane strain ($\varepsilon_{yy}$ for this set of axis definitions) in Figure 8b) from film to substrate suggesting coherent growth, consistent with prior atomic-resolution imaging [34]. The only exception is some red spots on the interface, which were previously found to be CoO precipitates [34]. There is, however, a slight compressive out-of-plane strain as seen in Figure 8a), as would be expected with LCMO having smaller lattice parameters than STO. Specifically, the out-of-plane spacing would be $4.480$Å for $\{011\}_M$ if totally unstrained. But the elastic strain imposed in-plane on the film from the substrate in this configuration (Figure 1b) would be tensile of $+0.26\%$ (averaging the three in-plane vector lengths compared to STO) which would cause an additional out-of-plane contraction by Poisson's effect. If we assume the same Poisson's ratio $\upsilon = 1$ model, as above, then this gives a total out-of-plane strain (referenced to substrate) of $\varepsilon_{xx} = -0.90\%$, which is the right order of magnitude for what is seen in the $\varepsilon_{xx}$ map. If there was constraint along the viewing direction too, the full Poisson's ratio calculation would come into effect and the $\varepsilon_{xx}$ would be higher.

As seen in Figure 8c) the film again deviates from 90° angles between the vectors, as expected for this structure, and these small deviations are in the range expected from Table 1. There is a noticeable switch in lattice rotation between left of and right of the right edge of the notched domain in the centre (Figure 8d).

Putting together the **d**-direction mapping of Figures 5 and 6 together with the strain maps of Figures 7 and 8, it turns out that these films are like many oxide thin films grown before, in that the dominant effect on domain selection is strain (for instance MacLaren *et al*. [2]). It is rare that this is not the case (a recent example was published where strain was small and oxygen octahedral coupling played a significant role [51]). However, unlike most previous studies, where simple diffraction in the ZOLZ or contrast in atomic resolution imaging were able to resolve the domain orientations and link them to film-substrate strain, this case gave no hints in atomic resolution STEM imaging [34] and little unambiguous information in the ZOLZ of diffraction patterns as to domain orientations. Only the occasional area oriented along $[100]_M$ gives an unambiguous low-angle diffraction pattern, and that is just a minority of areas in the film grown on STO, and no areas at all for LSAT.

So, the question remains as to why growth on LSAT should give better Co-Mn ordering than growth on STO [34]. If we have $[100]$ in-plane, then the film plane is either $(011)$ or $(01\bar{1})$. These have a lattice spacing of 4.48 Å in the bulk[31] and that might be slightly lower still with the compressive out-of-plane strain of about -0.5-0.75% seen in this film. On the other hand, if we have $[010]$ in-plane, the film plane is either $(101)$ or $(10\bar{1})$, which have plane spacings of 4.51 or 4.50 Å in bulk. Using the LSAT as a reference, then the out-of-plane strain is between 1.0 and 1.5%, and that corresponds to actual out-of-plane lattice parameters of 4.51-4.54 Å. With the benefit of hindsight, this does accord with Figure 2 of Kleibeuker *et al.* [34] where the out-of-plane lattice parameter for LCMO on STO is clearly smaller than that for LSAT. This difference might not sound much, but might make all the difference in atomic ordering. To get perfect atomic ordering, we need alternate layers of Mn and Co filling the B sites. Whilst $Mn^{4+}$ is relatively small, $Co^{2+}$ is relatively large (Bull *et al.*[52] found Mn-O distances of 1.89 Å and Co-O distances of 1.99 Å in ordered LCMO). Thus, the promotion of a crystal orientation that allows bigger spaces out-of-plane, and then increases the size of those spaces further with a little tensile strain in the same direction is going to be really favourable for creating layers that have enough space to fit the Co atoms in in a complete layer.

## E. General discussion

It is particularly enlightening that the presence or absence of the low angle FOLZ ring and the precise details of both the shift of diffraction spots in this ring and the azimuthal intensity variation around this are so rich in information. Of course, a similar richness of information not seen in the ZOLZ was found by Qiao to diagnose the presence or absence of charge density waves [27]. It is therefore interesting to wonder if there were ambiguities of interpretation in previous studies which were not resolved because of not having access to this information. For example, Woodward and Reaney [53] produced an excellent survey of the appearance of different families of superlattice spots in perovskites with different octahedral tilt systems, but unambiguous determination always depended on extensive tilting because it only used the ZOLZ. Heavy tilting is both time consuming even when possible, and impractical in thin films with small domains where overlaps of neighbouring domains will be a problem at higher tilts. In contrast to this, the new 4DSTEM methodology accomplishes this 3D crystallographic characterisation over a significant area of film in a single scan and with processing that can readily be automated now it has been demonstrated. As such, this can become a very powerful tool for the characterisation of complex oxide heterostructures and thereby invaluable in the development of oxide based functional devices.

It is noted that this work is effectively determining the full 3-dimensional crystal orientation from a nanobeam electron diffraction pattern, which is something that has been widely performed in recent years by matching experimental patterns to pre-computed banks of possible patterns for that crystal structure [24,54,55]. However, in contrast to such methods, this only chooses from a small set of possibilities. Whilst template-matching can be tweaked to only allow a smaller set of candidate patterns into the template bank (most easily to a limited angular range for materials known to have strong crystallographic texture), this picks possibilities that may lie far apart in any standard triangle, but which are crystallographically plausible in this context. Moreover, the critical choice is based on reflections in the FOLZ, which are typically weak and likely have a low weight in deciding the best fit. As such, using template-matching approaches would be both slow (hours for this, compared to DDF imaging, which takes about a second per calculation of an image on the server used in our work) and probably have a lot more uncertainty and noise in results. Additionally, by mapping the azimuthal intensity distribution, we see hints of rotation of the **d**-direction for La site positional modulations away from being perfectly parallel to the crystallographic **b**-axis. Incorporating crystal structure changes into crystal orientation mapping is hard, although strain measurements have been included in py4dstem [24]. Finally, standard crystallographic orientation mapping produces a set of Euler angles, from which Inverse Pole Figure maps can be produced. Determination of the orientation of a specific crystallographic direction in 3D space requires additional processing, which in the experience of one of the present authors, can reveal ambiguities in the indexing that are not visible in the IPF maps [56].

Additionally, an approach like the present one allows more flexibility to image changes in atom movements in a crystal as a function of position, even if care is needed with interpretation, via comparison to simulation. Such methods could be more widely applicable to thin film epitaxial growth of lower-symmetry oxides on perovskite substrates. Similarly, this could be applied to the characterisation of lower-symmetry oxides used as "single-crystal" substrates for epitaxial growth (such as $DyScO_3$) and how any domain structure in the substrate affects the film. Furthermore, similar effects where a single crystal produces a range of "domains" or "twin variants" or Widmanstätten patterns as a result of some crystallographic transformation with a clear parent-child orientation relationship may also result in a small list of possible patterns that could be quickly distinguished by Digital Dark Field imaging, and where use of the FOLZ or HOLZ could give additional certainty about the correctness of indexing, and thereby allow more unambiguous reconstruction of the crystallography in 3D.

# IV. Conclusions

We have performed Digital Dark Field Imaging of thin films of $La_2CoMnO_6$ films grown by pulsed laser deposition on $SrTiO_3$ and LSAT, using sets of superlattice reflections in the zeroeth, first, and second order Laue zones. The 3-dimensional information in the higher order Laue zones is much more specific and informative than just using the zeroeth order Laue zone alone, as has been standard practice in most dark field imaging to date. The azimuthal distribution of intensity in the first order Laue zone could also be determined. By combining the information in the different Laue zones, and using simulations of different diffraction patterns, it was possible to determine the 3D orientation of the La displacements in the crystal structure in a simple automated way. Thus, domain structures and crystal axes of the monoclinic structure were visualised in terms of 3D orientations of the **d**-directions, which should be parallel or close to the crystallographic **b**-axis. This revealed a huge difference between STO, in which case the **d**-directions, and therefore the **b**-axis preferred to sit out of the film plane (approximately 55° to the film plane) and LSAT, in which the **d**-directions, and thus the **b**-axis preferred to sit in the film plane. This was explained on the basis of the switch from tensile strain on STO to compressive strain in LSAT. Moreover, this switch to growth on a substrate that gives compressive in-pane strain creates a larger out-of-plane lattice spacing on LSAT, giving more room to fit in the larger $CoO_6$ octahedra, and thus promoting stronger Co-Mn ordering, which ultimately gives better properties. More generally, the methods herein would allow similar 3D characterisation of crystal orientations in other heteroepitaxial functional oxide systems where domain structures may play a strong role in driving the properties and performance.

## Acknowledgements

This work was only possible as a result of what was started by EPSRC funding of the grant "Fast Pixel Detectors: a paradigm shift in STEM imaging" (EP/M009963/1), and funded the computer server used in the calculations. Prof. Judith L. MacManus Driscoll and Dr. Josée Kleibeuker at the University of Cambridge are thanked for the provision of the films used in this work. IM is grateful to Prof Colin Ophus of Stanford University for helpful discussions, especially as regards making the dynamical simulations used in Figure 2.

# The structural effects of (111) growth of $La_2CoMnO_6$ on $SrTiO_3$ and LSAT – new insights from 3D crystallographic characterisation with 4D-STEM and Digital Dark Field imaging - Supplemental Materials

Ian MacLaren[1], Andrew T. Fraser[1], Matthew R. Lipsett[1]

1. School of Physics and Astronomy, University of Glasgow, Glasgow G12 8QQ, UK

## Effect of thickness and precession on intensity distribution in the FOLZ and determining **b**-axis orientation

Figure S1 shows a set of 6 simulations of $[111]_m$ diffraction patterns for thicknesses from 20 nm to 120 nm using the conditions used in the paper (i.e. with 0.1° precession on). The integrated intensity for each quadrant is calculated for each and ranked in order with highlights in red (most intense), yellow, green and blue (least intense) being used to show which is which. The most obvious effect of thickness is on the total amount of intensity in the diffraction spots, and this declines with thickness for obvious reasons of increased diffuse and inelastic scattering. No change in the ordering of quadrant is seen with thickness in this case.

Figure S2 shows the same simulation set, but this time with precession turned off. In one diffraction pattern, for 100 nm thickness the top-left, bottom-right is inverted. This emphasises that dynamical effects can change ordering and using the brightest quadrant to determine **b**-axis orientation using scanning electron nanodiffraction without precession is not robust. As might be expected, precession, even at tiny angles of 0.1° is doing enough to suppress dynamical effects to make the contrast more consistent and interpretable.

However, the geometry of the patterns is unaffected by thickness, as would also be expected. The peak azimuth is always in either the top-left or bottom-right quadrant, matching the 2D projected direction of the **b**-axis. And the shift of spots in the FOLZ with respect to ZOLZ is consistent throughout at ¾,0. This then provides a consistent determination of **b**-axis orientation in 3D, even if dynamical effects affect the intensities. A ¾,0 shift always correlates with the upwards tilt of **b** in the lower half of the figure. Putting this together with the intensity modulation telling us that the direction has to be lower right or upper left quadrant, then means that lower right is the direction in which the **b** axis comes up out of the page. The same works for all $\langle 111 \rangle_m$ and $\langle 11\bar{1} \rangle_m$ directions in whichever in-plane orientation is shown.

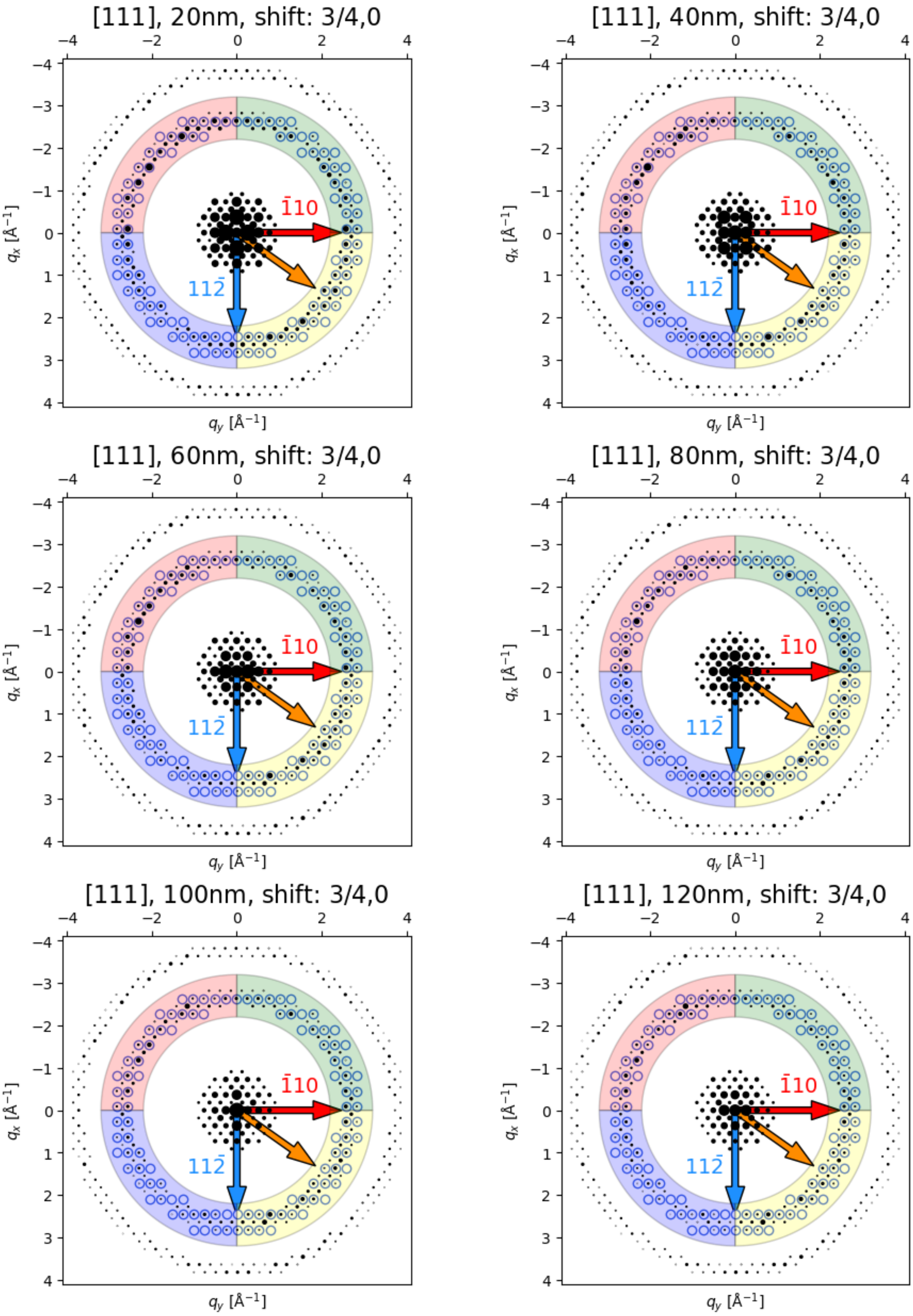

Figure S1: Dynamical simulations of the same diffraction pattern for six different thicknesses with precession = 0.1°. The quadrants of the FOLZ are highlighted in order of summed intensity: red (most intense), yellow, green, blue (least intense). The direction in which the **b**-axis of the structure points upwards out of the page is shown with an orange arrow. The brightest quadrant of the FOLZ is highlighted in pink.

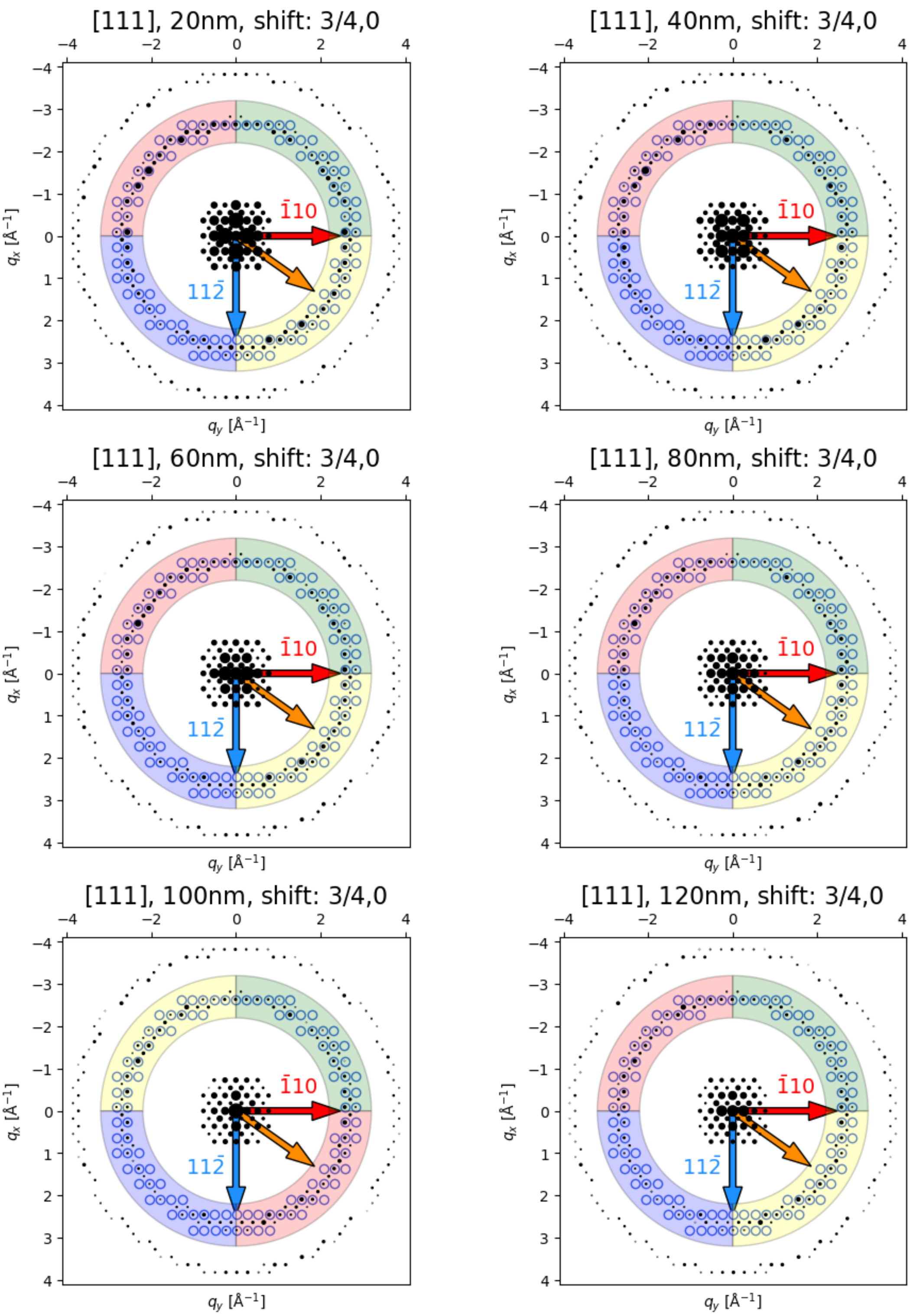

Figure S2: Dynamical simulations of the same diffraction pattern for six different thicknesses with precession off. The quadrants of the FOLZ are highlighted in order of summed intensity: red (most intense), yellow, green, blue (least intense). The direction in which the **b**-axis of the structure points upwards out of the page is shown with an orange arrow. The brightest quadrant of the FOLZ is highlighted in pink.

## Signal levels, SNR and SBR in Digital Dark Field Imaging

Figure S3 shows the images of Figure 3 with calculations of average and standard deviation signals for each box, together with calculations of Signal to Noise Ratio (SNR) and Signal to Background Ratio (SBR). Though the average counts per pixel varies by a factor of ~ 250 from ~60,000 to ~ 233, imaging is useful in all cases with recognisable features. SBR is characteristically high since the background is at or close to zero in most cases. SNR does trend down with reduced counts, but not catastrophically and it is perhaps only in the last image where one wonders if single pixel variations could be discerned reliably (since this is below the Rose criterion).

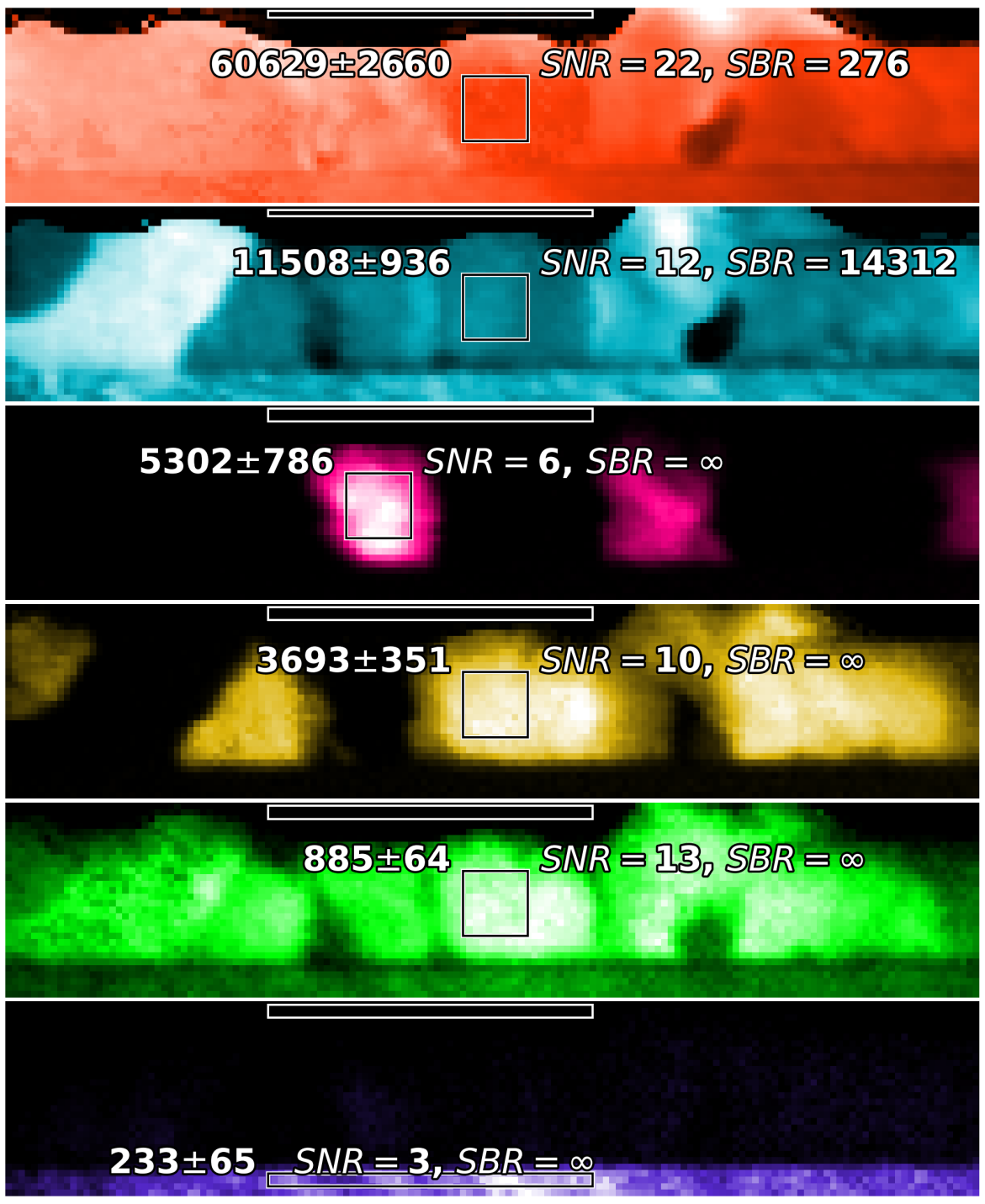

Figure S3: The images of Figure 3 reprinted with average and standard deviations of the signals in the black boxes, together with SNR, and SBR calculated with reference to the white boxes at the top of each image.